\documentclass[12pt]{article}

\usepackage[utf8]{inputenc}
\usepackage{amsmath, amsthm, amssymb, amsfonts}
\usepackage{import}
\usepackage[numbers, comma]{natbib}
\usepackage{fullpage}
\usepackage{authblk}
\usepackage[bibliography=common]{apxproof}
\usepackage{import}
\usepackage{mathtools}
\usepackage{tikz}
\usepackage{float}
\usepackage{enumitem}

\usetikzlibrary{calc,patterns,angles,shapes, positioning, intersections, quotes}

\newtheorem{theorem}{Theorem}
\newtheoremrep{thm}{Theorem}
\newtheoremrep{lem}{Lemma}

\usepackage{graphicx}
\newtheorem{corollary}[theorem]{Corollary}

\theoremstyle{definition}

\newtheorem{assumption}{Assumption}
\usepackage[colorlinks=false,pagebackref=true, hidelinks]{hyperref}

\allowdisplaybreaks

\date{}

\begin{document}
\title{Optimal tie-breaking rules\thanks{We are grateful to Arunava Sen, Federico Echenique, Wade Hann-Caruthers, the Editor Carmen Bevia, and referees for this journal, as well as referees and participants at the 9th Annual Conference on Contests: Theory and Evidence (2023) for helpful comments and suggestions.}
}

\author{Sumit Goel\thanks{California Institute of Technology; \href{mailto:sumitgoel58@gmail.com}{sumitgoel58@gmail.com}; 0000-0003-3266-9035} \quad Amit Goyal\thanks{Indian Statistical Institute Delhi; \href{mailto:amit.kr.goyal@gmail.com}{amit.kr.goyal@gmail.com}}}

\maketitle        

\begin{abstract}
We consider two-player contests with the possibility of ties and study the effect of different tie-breaking rules on effort. For ratio-form and difference-form contests that admit pure-strategy Nash equilibrium, we find that the effort of both players is monotone decreasing in the probability that ties are broken in favor of the stronger player. Thus, the effort-maximizing tie-breaking rule commits to breaking ties in favor of the weaker agent. With symmetric agents, we find that the equilibrium is generally symmetric and independent of the tie-breaking rule. We also study the design of random tie-breaking rules that are ex-ante fair and identify sufficient conditions under which breaking ties before the contest actually leads to greater expected effort than the more commonly observed practice of breaking ties after the contest.
\end{abstract}
\section{Introduction}

Contests are situations in which agents exert costly effort to win one or more prizes. Examples of such competitive situations include sporting contests, promotional tournaments, political contests, R\&D races, etc. In many of these situations, it is often the case that there is no outright winner and the contest ends in a draw or a tie. Moreover, a draw may not be an acceptable outcome for the designer. For instance, in sports competitions such as cricket, chess, and soccer, a significant fraction of the games end in a draw. But if these games happen to be knockout games of a world event, the designer must determine a single winner. Many different tie-breaking rules have been used to determine a winner in such situations. The result might be decided by chance\footnote{The outcome of many elections that ended with ties have been determined via a coin toss (e.x. \href{https://www.bbc.com/news/world-us-canada-63599832}{Kentucky city mayor race 2022}). Even in sporting competitions where ties are broken by another short duration contest, like the super over in cricket or the penalty shootout in soccer, it can be argued that the outcome is almost a random draw as there is relatively little scope for skill or effort to have an impact.} or it might be pre-determined based on some personal attributes like age, sex, height, weight, or status (incumbent or challenger)\footnote{In \href{https://thebarbellspin.com/weightlifting/bodyweight-tiebreaker-to-be-removed-per-iwf-executive-board/}{weightlifting contests}, ties were resolved, until 2016, in favor of the lighter athlete. In boxing, if a championship bout ends in a draw, the champion usually retains the title. In the 1999 Cricket World Cup, the \href{https://en.wikipedia.org/wiki/1999_Cricket_World_Cup_2nd_semi-final}{tied semi-final match between Australia and South Africa} went to Australia because they had defeated South Africa earlier in the tournament.}. In this paper, we consider contests between two agents and focus on understanding how different tie-breaking rules compare in terms of the effort they induce. \\

We study the effect of tie-breaking rules for three contest environments that differ in how the effort exerted by the players determine the distribution over contest outcomes (player 1 wins, player 2 wins, or tie). We study ratio-form contests (\citet{tullock2001efficient, baik2004two, ewerhart2017revenue, wang2010optimal, nti1999rent, nti2004maximum}), difference-form contests (\citet{hirshleifer1989conflict, baik1998difference, skaperdas1996contest, bevia2015relative, che2000difference}), and lastly, we discuss examples of concave contests (\citet{blavatskyy2010contest, fu2020optimal}). While the literature has typically assumed only two possible outcomes\footnote{Surveys of this literature can be found in \citet{garfinkel2007economics, jia2013contest, corchon2018contest, chowdhury2023heterogeneity, mealem2016discrimination}}, we will consider generalizations of these contests that allow for the possibility of ties (\citet{blavatskyy2010contest, jia2012contests, vesperoni2019contests}). A feature of some of these generalizations is that the probability of tie increases as the efforts come closer, and in particular, it is maximized when both agents exert equal efforts. Motivated by this, with ratio-form and difference-form contests, we assume that the probability of a tie increases as the contest becomes more equal (ratio of efforts goes to 1 or difference goes to 0). In addition, we make assumptions that ensure existence and uniqueness of pure-strategy Nash equilibrium. \\

We make two primary contributions. First, we find that when the two agents differ in their value from winning, their effort is decreasing in the probability that the ties are broken in favor of the stronger player. Thus, an effort-maximizing contest designer would prefer to commit to breaking ties in favor of the weaker player. In this way, our results lend support to the idea of leveling the playing field and increasing the competitive balance of a contest to increase effort. Second, we find that with symmetric agents, the equilibrium is generally symmetric and independent of the tie-breaking rule. We also discuss the design of random unbiased tie-breaking rules and identify conditions under which breaking ties before the contest would lead to greater effort than the standard practice of breaking them after the contest has ended in a tie. For all our results, we identify parametric classes of contests that satisfy our assumptions and discuss the application of our results to them. \\

There is a growing literature studying the effect of introducing ties on equilibrium effort in contests. For concave contests with ties, introduced by \citet{loury1979market} and axiomatized by \citet{blavatskyy2010contest}, the possibility of a tie has been shown to reduce total equilibrium effort (\citet{nti1997comparative, deng2018incentives, li2023optimally}) though it may increase winner's expected effort (\citet{deng2018incentives}) or even total effort in a winner-pay setting (\citet{minchuk2022winner}).  In all-pay auctions, finite strategy spaces and bid gaps have been used to study the effect of introucing ties (\citet{eden2006optimal, cohen2007contests,gelder2019all}). Other related work has illustrated the merits of introducing draws in different contests (\citet{nalebuff1983prizes,imhof2014tournaments,  imhof2016ex, chang2023analysis}). In comparison to this literature, a tie is a natural outcome in our contests and the prize is awarded irrespective of the contest outcome. \\


The paper contributes to the literature studying the effect of draw prizes on effort. Most of the work in this domain assumes symmetric agents and studies the effect of a common draw prize on effort. For a generalization of ratio-form contests to allow for ties, \citet{vesperoni2019contests} find that there is a unique symmetric equilibrium that does not depend on the common draw prize\footnote{They also find conditions under which the equilibrium effort is greater with the possibility of draw as compared to without it, even though the draw prize may be $0$.}. In ratio-form contests, \citet{jia2012contests} shows that there is always a unique equilibrium that is symmetric even though the contest might be biased\footnote{This is referred to as the homogeneity paradox as one would probably expect that players put in different levels of effort if the contest favors one player over another.}. They illustrate how introducing ties with no prizes can lead to equilibrium that is no longer symmetric. In comparison to this literature, our model allows for asymmetric agents and can be interpreted as studying the effects of awarding complementary draw prizes. We find that for both \citet{vesperoni2019contests} and \citet{jia2012contests} contests,  the equilibrium effort is decreasing in the probability that a tied contest is awarded to the stronger player. In a similar spirit, \citet{szech2015tie} finds that an asymmetric tie-breaking rule that favors the weaker player increases expected bids in an all-pay auction environment.\\

The paper also contributes to the literature studying the effect of leveling the playing field on effort in contests with heterogeneous agents. There are many different mechanisms that have been studied including multiplicative biases and additive headstarts\footnote{Surveys of this literature can be found in \citet{mealem2016discrimination,chowdhury2023heterogeneity}.}. In short, it has been found that if the designer cares about increasing effort and can optimally choose multiplicative biases, it cannot gain from being able to further choose additive biases or introduce draws (\citet{fu2020optimal, franke2018optimal, li2023optimally}). We believe ours is the first to consider this idea of leveling the playing field in the context of tie-breaking rules. We note here that for some contests (like the class of concave contests from \citet{fu2020optimal}), the choice of a tie-breaking rule is essentially equivalent to defining different head-starts for the two players. However, the tie-breaking rule imposes constraints on the set of feasible head-starts (such as the sum of head-starts must be a constant) and so the problem of finding the optimal tie-breaking rule in these contests is still different from the problem of finding optimal headstarts as considered in these papers.   \\

The paper proceeds as follows. In section 2, we present our model of a two-player contest with ties. In section 3 and 4, we study ratio-form and difference-form contests respectively. In section 5, we discuss some other contest examples and study the design of random unbiased tie-breaking rules. Section 6 concludes. Most of the proofs are relegated to the appendix.

\section{Model}

There are two risk-neutral players competing in a contest. The contest has three possible outcomes: player 1 wins, player 2 wins, or it may be a tie. The distribution over the three outcomes is determined by the efforts $x_1 \geq 0, x_2 \geq 0$ exerted by the two players. We let $p_i(x_1,x_2)$ denote the probability that player $i$ wins and $p_0(x_1,x_2)$ denote the probability that the contest ends in a tie. Thus, for any $x_1,x_2 \geq 0$, $p_1+p_2+p_0=1$.  In case of a tie, the designer awards the contest to player 1 with probability $q \in [0,1]$ and awards it to player $2$ with remaining probability $1-q$. We refer to $q$ as the \textit{tie-breaking rule}. Note that $q$ is independent of the effort exerted by the agents. Given the tie-breaking rule $q$, player $1$ eventually wins the contest with probability $P_1=p_1+qp_0$ while player $2$ eventually wins with probability $P_2=p_2+(1-q)p_0$. The value of player $i\in \{1,2\}$ from winning the contest is $V_i$, where we assume $V_1 \geq V_2 >0$. The cost of exerting effort $x_i$ for player $i$ is given by $c(x_i)$ where $c:\mathbb{R}_+\to \mathbb{R}_+$ is a cost function. We will assume either linear $c(x)=x$ or quadratic costs $c(x)=x^2$ in our analysis.

Given a contest $C=(V_1,V_2,p_1(), p_2(), q, c())$, player $1$'s payoff under profile $(x_1,x_2)$ is
$$\Pi_1(x_1,x_2)=V_1(p_1(x_1,x_2)+qp_0(x_1,x_2))-c(x_1),$$ and that of player $2$ is
$$\Pi_2(x_1,x_2)=V_2(p_2(x_1,x_2)+(1-q)p_0(x_1,x_2))-c(x_2).$$
An effort profile $(x_1^*, x_2^*)$ is a pure-strategy Nash equilibrium if
$$\Pi_1(x_1^*, x_2^*) \geq \Pi_1(x_1, x_2^*) \text{ for all } x_1 \in \mathbb{R}_+  \text{ and } \Pi_2(x_1^*, x_2^*) \geq \Pi_2(x_1^*, x_2) \text{ for all } x_2 \in \mathbb{R}_+.$$

We will impose conditions on $C$ that guarantee the existence and uniqueness of a pure-strategy Nash equilibrium, characterized by the first-order conditions. We note here the two first-order conditions. The first-order condition for player $1$ is
\begin{equation}
\label{foc1}
    \dfrac{\partial \Pi_1}{\partial x_1}=0 \implies V_1 \left(\dfrac{\partial p_1}{\partial x_1}+q \dfrac{\partial p_0}{\partial x_1}\right)=c'(x_1),
\end{equation}

and that for player $2$ is

\begin{equation}
\label{foc2}
    \dfrac{\partial \Pi_2}{\partial x_2}=0 \implies V_2 \left(\dfrac{\partial p_2}{\partial x_2}+(1-q) \dfrac{\partial p_0}{\partial x_2}\right)=c'(x_2).
\end{equation}

Given $V_1,V_2, p_1(), p_2(), c()$, we consider the designer's problem of choosing a tie-breaking rule $q \in [0,1]$ so as to maximize the total effort in the pure-strategy Nash equilibrium\footnote{We note here that for some of our results, we only need the objective to be an increasing function of the efforts.}. Formally, the designer's problem is $$\max_{q \in [0,1]} x_1(q)+x_2(q)$$
where $x_i(q)$ refers to the equilibrium effort level of player $i$ under the contest given by $C=(V_1,V_2,p_1(),p_2(), q, c())$. Going forward, we will study and solve the designer's problem for different classes of contest games.



\section{Ratio-form contest success functions}

In this section, we consider instances of our model where the distribution over the three outcomes depends only on  the ratio of the efforts $\theta=\dfrac{x_1}{x_2}$ of the two players. Formally, we assume that there exist twice-differentiable functions $p:\mathbb{R}_+\to [0,1]$ and $p_0:\mathbb{R}_+\to [0,1]$ such that
\begin{itemize}
    \item $p_1(x_1, x_2)=p(\theta)$ and $p_2(x_1,x_2)=p(\frac{1}{\theta})$,
    \item $p_0(x_1, x_2)=p_0(\theta)$ and $p_0(\theta)=p_0(\frac{1}{\theta})$ for all $\theta \in \mathbb{R}_+$,
    \item $c(x)=x$.
\end{itemize}

We'll make the following assumption on these functions.

\begin{assumption}
\label{ass:existence}
Let $z_q(\theta)=p(\theta)+qp_0(\theta)$.
\begin{itemize}
\item For any $q\in [0,1]$,  $z_q^{\prime}(\theta)>0, z_q^{\prime \prime}(\theta)<0$, and $2 z_q^{\prime}(\theta)+\theta z_q^{\prime \prime}(\theta)>0$, for all $\theta$ in $R_{+}$.
\item For any $q\in [0,1]$, $z_q(0)=0$ and $\lim_{\theta \to \infty} z_q(\theta)=1$ 
\end{itemize}
\end{assumption}

Note that $z_q(\theta)$ denotes the probability that player $1$ eventually wins the prize and $1-z_q(\theta)$ is the probability that player $2$ wins eventually wins the prize. The first part of the assumption then says that a player's probability of winning is increasing in its own effort, decreasing in the effort of the other player, and in addition, it is increasing at a decreasing rate in one's own effort. The second part of the assumption says that if a player exerts $0$ effort while the other player exerts a positive effort level, its probability of winning is $0$ irrespective of the tie-breaking rule $q$. 

The next assumption we make is on the probability of ties $p_0(\theta)$ and is motivated by the idea that the probability of a tie increases as the contest becomes more closely contested. 

\begin{assumption}
\label{ass:tie}
The probability of tie $p_0(\theta)$ is increasing for $\theta \in (0,1]$ and decreasing for $\theta \in [1, \infty)$. 
\end{assumption}

We first provide a characterization of the pure-strategy Nash equilibria of the ratio-form contest game with ties.

\begin{thmrep}
\label{baik}
Consider a ratio-form contest satisfying Assumption \ref{ass:existence}. The contest has a unique pure-strategy Nash equilibrium defined by 
$$x_1^*=V_1 \beta z_q'(\beta) \text{ and } x_2^*= V_2 \beta  z_q'(\beta), $$
where $\beta=\frac{V_1}{V_2}\geq 1$.
\end{thmrep}

\begin{proof}

In a ratio-form contest $C=(V_1,V_2,p_1(),p_2(),q,c())$, there exist functions $p$ and $p_0$ so that we can rewrite players payoff under profile $x_1,x_2$ as $$\Pi_1=V_1z_q(\theta)-x_1 \text{ and } \Pi_2=V_2(1-z_q(\theta))-x_2 $$
where $\theta=\dfrac{x_1}{x_2}$ and $z_q(\theta)=p(\theta)+qp_0(\theta)$.

Since the prize is valued positively by both players and the probability of winning the contest with $x_i=0, x_j>0$ is $0$ for player $i$ and $1$ for player $j$, any pure-strategy Nash equilibrium involves positive effort levels by both players.

Now, observe that  
$$\dfrac{\partial \Pi_1}{\partial x_1}=\frac{V_1}{x_2}z_q'(\theta)-1 \text{ and } \dfrac{\partial \Pi_2}{\partial x_2}=\frac{x_1V_2}{x_2^2}z_q'(\theta)-1 $$

and 

$$\dfrac{\partial^2 \Pi_1}{\partial x_1^2}=\frac{V_1}{x^2_2}(z_q''(\theta))<0\text{ and } \dfrac{\partial^2 \Pi_2}{\partial x_2^2}=-\frac{x_1V_2}{x_2^3}\left(2z_q'(\theta)+\theta z_q''(\theta)\right)<0 .$$

From Assumption \ref{ass:existence}, $\Pi_i$ is concave in $x_i$ for any $x_j$. Since any pure-strategy Nash equilibrium must satisfy the first-order conditions
$$V_1z_q'(\theta)=x_2  \text{ and } x_1V_2 z_q'(\theta)=x_2^2,$$
concavity of payoffs implies that the second-order conditions are satisfied at any solution to the first-order conditions. 
Observe that any solution to the two first-order conditions must satisfy $$\dfrac{x_1}{x_2}=\dfrac{V_1}{V_2}=\beta.$$
But this condition uniquely pins down the solution to the first-order conditions: $$x_1^*=V_1\beta z_q'(\beta)   \text{ and }  x_2^*=V_2 \beta z_q'(\beta).$$


\end{proof}


To prove the result, we show that there  is a unique solution to the first-order conditions \ref{foc1} and \ref{foc2} for the contest game defined by $x_1^*$ and $x_2^*$. The concavity of the payoffs implied by Assumption \ref{ass:existence} means that the second-order conditions for maximization are satisfied\footnote{Note that Assumption \ref{ass:existence} implies that the players payoffs $\Pi_i$ are strictly concave in their actions $x_i$. Since the argument relies on there being a unique solution to the first-order conditions, we can relax Assumption \ref{ass:existence} so that payoffs are only required to be quasiconcave and the result would go through.}. We note here that the result also follows from the result of 
\citet{baik2004two} who characterized the pure-strategy Nash equilibrium of two player ratio-form contests satisfying Assumption \ref{ass:existence}.


The next result shows that the effort-maximizing tie-breaking rule breaks ties in favor of the weaker agent. 

\begin{thmrep}
\label{thm:ratio}
Consider a ratio-form contest satisfying Assumptions \ref{ass:existence} and \ref{ass:tie}.
\begin{itemize}
\item If $V_1>V_2$, $x_1^*(q)+x_2^*(q)$ is decreasing in $q$ and so the optimal tie-breaking rule is $q=0$. 
\item If $V_1=V_2$, $x_1^*(q)+x_2^*(q)$ is independent of $q$. 
\end{itemize}
\end{thmrep}
\begin{proof}[Proof of Theorem \ref{thm:ratio}]
From Theorem \ref{baik}, the total effort in equilibrium is given by $$R(q)=(V_1+V_2)\beta(p'(\beta)+qp_0'(\beta)).$$ Observe that the total effort is linear in $q$ with the coefficient $(V_1+V_2)\beta p_0'(\beta)$ and the sign of the coefficient is determined by $p_0'(\beta)$.  When $\beta=\frac{V_1}{V_2}>1$ Assumption \ref{ass:tie} implies that $p_0'(\beta)<0$ and it follows that the the total effort is decreasing in $q$. So the optimal tie-breaking rule breaks ties in favor of the weaker player $(q=0)$. 

Similarly, when $\beta=\frac{V_1}{V_2}=1$ Assumption \ref{ass:tie} implies that $p_0'(\beta)=0$ and it follows that the equilibrium effort, and thus, the total effort, is independent of $q$. So the choice of the tie-breaking rule does not matter in this case.

\end{proof}

First consider the case where the agents are symmetric so that $\beta=\frac{V_1}{V_2}=1$. Observe that in this case, both agents exert equal effort, irrespective of the bias introduced by the tie-breaking rule $q$. Intuitively, this is because at any symmetric profile, a player's effort has zero marginal impact on the probability of a tie, and therefore, the tie-breaking rule $q$ does not affect the symmetric equilibrium.  With asymmetric agents, we again have that the stronger agent puts in greater effort, irrespective of the bias due to the tie-breaking rule $q$. In particular, the marginal impact of $q$ on player $i$'s effort is simply $V_i\beta p_0'(\beta)$. Since $\beta>1$, $p_0'(\beta)<0$ and it follows that the effort of both agents goes down as $q$ increases. Thus, even if the designer had a more general objective function that was increasing in the effort of both agents, it would still be optimal for the designer to bias the tie-breaking rule completely in favor of the weaker player with $q=0$. Formally, the result follows from the fact that the total equilibrium effort is linear in $q$ with the coefficient $(V_1+V_2)\beta p_0'(\beta)$. The full proof is in the appendix.\\

The Tullock contests, defined by $p_i(x_i,x_{-i})=\frac{x_i^r}{x_i^r+x_{-i}^r}$ for $r\in (0, \infty)$, is an important class of ratio-form contest success functions that have been widely studied in the literature. There has been significant work in generalizing the Tullock contests to allow for the possibility of ties. We will consider three such generalizations. For two of these generalizations, the contest success functions take the ratio form and we discuss how our results apply to them next. The third generalization by \citet{blavatskyy2010contest} is discussed later.\\

The first generalization we consider is given by $p_i(x_i,x_j)=\frac{x_i^{rk}}{(x_i^r+x_j^r)^k}$ with $k\geq 1$ so that the probability of ties is $p_0(x_i, x_j)=\frac{(x_i^r+x_j^r)^k-x_i^{rk}-x_j^{rk}}{(x_i^r+x_j^r)^k}$. This was proposed by \citet{vesperoni2019contests}.  In this case, we obtain the following lemma.\\

\begin{lemrep}
\label{lem:vesper}
 Suppose $C$ is a ratio-form contest with $p_i(x_i,x_j)=\frac{x_i^{rk}}{(x_i^r+x_j^r)^k}$ where $k\geq 1$. If $rk \leq 1$, then for any $q\in [0,1]$, the contest has a unique pure-strategy Nash equilibrium defined by
 $$x_1^*=V_1 rk \dfrac{\beta^{rk}+q(\beta^{r}-\beta^{rk})}{(1+\beta^r)^{k+1}}  \text{ and }  x_2^*= V_2 rk \dfrac{\beta^{rk}+q(\beta^{r}-\beta^{rk})}{(1+\beta^r)^{k+1}},$$ where $\beta=\frac{V_1}{V_2}$. 
\end{lemrep}
\begin{proof}
To prove the result, we define the functions $p(\theta), p_0(\theta)$ and show that they satisfy Assumptions \ref{ass:existence}, \ref{ass:tie} when $rk \leq 1$ with $r \leq 1$, $k\geq 1$. 

For the given contest, we can define the functions $$p(\theta)=\dfrac{\theta^{rk}}{(1+\theta^r)^k}, \quad p_0(\theta)=\dfrac{(1+\theta^r)^k-1-\theta^{rk}}{(1+\theta^r)^k}$$ 

so that $$z_q(\theta)=\dfrac{\theta^{rk}+q((1+\theta^r)^k-1-\theta^{rk})}{(1+\theta^r)^k}.$$

Let us first identify conditions under which Assumption \ref{ass:existence} is satisfied.

To do so, we have that

\begin{align*}
z_q'(\theta)&=\dfrac{(1+\theta^r)^{k}(rk\theta^{rk-1}(1-q)+qkr(1+\theta^r)^{k-1}\theta^{r-1})-(\theta^{rk}+q((1+\theta^r)^k-1-\theta^{rk}))(kr(1+\theta^r)^{k-1}\theta^{r-1})}{(1+\theta^r)^{2k}}\\
&=kr\theta^{r-1}\dfrac{(\theta^{rk-r}(1-q))+q}{(1+\theta^r)^{k+1}}\\
&=kr \dfrac{\theta^{rk-1}(1-q)+q\theta^{r-1}}{(1+\theta^r)^{k+1}}.\\
\end{align*}

Also, 

\begin{align*}
z_q''(\theta)&=kr\theta^{r-2}\dfrac{q(r-1)+\theta^{rk-r}(rk-1)(1-q)-\theta^{rk}(1+r)(1-q)-\theta^r q(1+kr)}{(1+\theta^r)^{k+2}}\\
&=kr\theta^{r-2}\dfrac{q(r-1-\theta^r(1+kr))+(1-q)\theta^{rk-r}(rk-1-\theta^r(1+r))}{(1+\theta^r)^{k+2}}.
\end{align*}

Lastly, 

\begin{align*}
    2z_q'(\theta)+\theta z_q''(\theta)&=kr\theta^{r-1}\dfrac{q(r+1+\theta^r(1-kr))+(1-q)\theta^{rk-r}(rk+1+\theta^r(1-r))}{(1+\theta^r)^{k+2}}.
\end{align*}

From the above expressions, we have that $z_q'(\theta)>0$ for all $\theta>0$. The condition $rk\leq 1$ with $r\leq 1$ and $k\geq 1$ is sufficient for both $z_q''(\theta)<0$ and $2z_q'(\theta)+\theta z_q''(\theta)>0$. Thus, we have that if $rk \leq 1$, Assumption \ref{ass:existence} is satisfied.

Observe that
$$p_0'(\theta)=kr\dfrac{\theta^r-\theta^{rk}}{\theta(1+\theta^r)^{k+1}},$$
which is $\geq 0$ for $\theta\leq 1$ and $\leq 0$ for $\theta\geq 1$ since $k\geq  1$. Thus, Assumption \ref{ass:tie} is also satisfied.

\end{proof}

To prove the result, we define the functions $p(\theta), p_0(\theta)$ and show that they satisfy Assumptions \ref{ass:existence}, \ref{ass:tie} when $rk \leq 1$ with $r \leq 1$, $k\geq 1$. Note that the conditions in the lemma are sufficient but not necessary for the result. The full proof is in the appendix.  \\

The second generalization we consider is given by $p_i(x_i,x_j)=\frac{x_i^{r}}{x_i^r+kx_j^r}$ with $k \geq 1$ so that the probability of ties is $p_0(x_i, x_j)=1-\frac{x_i^{r}}{x_i^r+kx_j^r}-\frac{x_j^{r}}{kx_i^r+x_j^r}$. This was proposed by \citet{jia2012contests}.  In this case, we obtain the following lemma.

\begin{lemrep}
\label{lem:jia}
 Suppose $C$ is a ratio-form contest with $p_i(x_i,x_j)=\frac{x_i^{r}}{x_i^r+kx_j^r}$ where $k\geq 1$. If $r \leq 1$, then for any $q\in [0,1]$, the contest has a unique pure-strategy Nash equilibrium defined by
 $$x_1^*=V_1rk\beta^r \left(\frac{1}{(\beta^r+k)^2}+q\left(\frac{1}{(1+k\beta^r)^2}-\frac{1}{(\beta^r+k)^2}\right)\right)$$ and 
 $$x_2^*=V_2rk\beta^r \left(\frac{1}{(\beta^r+k)^2}+q\left(\frac{1}{(1+k\beta^r)^2}-\frac{1}{(\beta^r+k)^2}\right)\right),$$ where $\beta=\frac{V_1}{V_2}$. 
\end{lemrep}

\begin{proof}
To prove the result, we define the functions $p(\theta), p_0(\theta)$ and show that they satisfy Assumptions \ref{ass:existence}, \ref{ass:tie} when $r \leq 1$ . We further show that the condition in Theorem \ref{thm:ratio} is also satisfied when $r \leq 1$  which implies the result.

For the given contest, we can define the functions $$p(\theta)=\dfrac{\theta^r}{\theta^r+k},\quad p_0(\theta)=1-\dfrac{\theta^r}{\theta^r+k}-\dfrac{1}{k\theta^r+1}$$ 

so that $$z_q(\theta)=\dfrac{\theta^r(1+k \theta^r+q(k^2-1))}{(1+k\theta^r)(\theta^r+k)}.$$

Let us first identify conditions under which Assumption \ref{ass:existence} is satisfied.

To do so, we have that

\begin{align*}
z_q'(\theta)&=kr\theta^{r-1}\dfrac{(1-q)(1+k\theta^r)^2+q(\theta^r+k)^2}{(1+k\theta^r)^2(\theta^r+k)^2}.
\end{align*}

Also, 

\begin{align*}
z_q''(\theta)&=-kr\theta^{r-2}\left(2r\theta^r\left(\dfrac{1-q}{(\theta^r+k)^3}+\dfrac{qk}{(1+k\theta^2)^3}\right)+(1-r)\left(\dfrac{1-q}{(\theta^r+k)^2}+\dfrac{q}{(1+k\theta^2)^2}\right)\right).
\end{align*}

Lastly, 

\begin{align*}
    2z_q'(\theta)+\theta z_q''(\theta)&=kr\theta^{r-1}\left(\dfrac{(1-q)}{(\theta^r+k)^2}\left(1+r-\dfrac{2r\theta^r}{\theta^r+k}\right)+\dfrac{q}{(1+k\theta^r)^2}\left(1+r-\dfrac{2rk\theta^r}{1+k\theta^r}\right)\right).
\end{align*}

From the above expressions, we have that $z_q'(\theta)>0$ for all $\theta>0$. The condition $r\leq 1$ is sufficient for both $z_q''(\theta)<0$ and $2z_q'(\theta)+\theta z_q''(\theta)>0$. Thus, we have that if $r \leq 1$, Assumption \ref{ass:existence} is satisfied.

Observe that

\begin{align*}
p_0'(\theta)&=kr\theta^{r-1}\dfrac{(\theta^r+k)^2-(1+k\theta^r)^2}{(1+k\theta^r)^2(\theta^r+k)^2}\\
&=kr\theta^{r-1}\dfrac{(\theta^r+k+1+k\theta^r)(\theta^r+k-1-k\theta^r)}{(1+k\theta^r)^2(\theta^r+k)^2}\\
&=kr\theta^{r-1}\dfrac{(\theta^r+k+1+k\theta^r)(k-1)(1-\theta^r)}{(1+k\theta^r)^2(\theta^r+k)^2},
\end{align*}
which is $\geq 0$ for $\theta\leq 1$ and $\leq 0$ for $\theta\geq 1$ since $k\geq  1$. Thus, Assumption \ref{ass:tie} is also satisfied.






    
\end{proof}

Again, we define the functions $p(\theta), p_0(\theta)$ and show that Assumptions \ref{ass:existence} and \ref{ass:tie} are satisfied when $r \leq 1$. The full proof is in the appendix. Observe that in Lemmas \ref{lem:vesper} and \ref{lem:jia}, if we plug in $k=1$, we get back the equilibrium characterization for the Tullock contests without ties (\citet{nti1999rent}). Also, the total effort under the optimal tie-breaking rule $(q=0)$ in the \citet{vesperoni2019contests} contest is $(V_1+V_2)rk\frac{\beta^{rk}}{(1+\beta^r)^{k+1}}$ and that under the \citet{jia2012contests} contest is $(V_1+V_2)rk\frac{\beta^{r}}{(k+\beta^r)^{2}}$. Since both are increasing in $k$ at $k=1$, it follows that if the designer could induce ties by increasing $k$ and then commit to breaking them in favor of the weaker player, it would induce greater effort than in the standard Tullock contests.

\section{Difference-form contest success functions}

In this section, we consider instances of our model where the distribution over the three outcomes depends only on  the difference of the efforts $\theta=x_1-x_2$ of the two players. Formally, we assume that there exist twice-differentiable functions $p:\mathbb{R}\to [0,1]$ and $p_0:\mathbb{R}\to [0,1]$ such that
\begin{itemize}
    \item $p_1(x_1, x_2)=p(\theta)$ and $p_2(x_1,x_2)=p(-\theta)$,
    \item $p_0(x_1, x_2)=p_0(\theta)$ and $p_0(\theta)=p_0(-\theta)$ for all $\theta \in \mathbb{R}$,
    \item $c(x)=x^2/2$.
\end{itemize}

We'll make the following assumption on these functions.

\begin{assumption}
\label{ass:existence2}
Let $z_q(\theta)=p(\theta)+qp_0(\theta)$. For any $q\in [0,1]$,  $z_q^{\prime}(\theta)>0$ and $z_q''(\theta) \in [-\frac{1}{V_1},\frac{1}{V_1}]$.
\end{assumption}

Here again,  $z_q(\theta)$ denotes the probability that player $1$ eventually wins the prize and $1-z_q(\theta)$ is the probability that player $2$ wins eventually wins the prize. The assumption then says that a player's probability of winning is increasing in its own effort, decreasing in the effort of the other player. The second assumption ensures that the player's objective function is globally concave and a unique best response exists. 

The next assumption we make is on the probability of ties $p_0(\theta)$ and is again motivated by the idea that the probability of a tie increases as the contest becomes more closely contested. 

\begin{assumption}
\label{ass:tie2}
The probability of tie $p_0(\theta)$ is increasing for $\theta \in (-\infty,0]$ and decreasing for $\theta \in [0, \infty)$. 
\end{assumption}

Let us now characterize the pure-strategy Nash equilibria of this difference-form contest game with ties.

\begin{thmrep}
\label{thm:diff1}
Consider a difference-form contest satisfying Assumption \ref{ass:existence2}. The contest  has a unique pure-strategy Nash equilibrium defined by 

$$x_1^*=V_1z_q'(\beta(q))   \text{ and }  x_2^*=V_2z_q'(\beta(q)),$$
where $\beta(q)$ is the unique solution to the equation $\theta=(V_1-V_2)z_q'(\theta)$.
\end{thmrep}

\begin{proof}

In a difference-form contest $C=(V_1,V_2,p_1(),p_2(),q,c())$, there exist functions $p$ and $p_0$ so that we can rewrite players payoff under profile $x_1,x_2$ as $$\Pi_1=V_1z_q(\theta)-\frac{x_1^2}{2} \text{ and } \Pi_2=V_2(1-z_q(\theta))-\frac{x_2^2}{2} $$
where $\theta=x_1-x_2$ and $z_q(\theta)=p(\theta)+qp_0(\theta)$.


Now, observe that  
$$\dfrac{\partial \Pi_1}{\partial x_1}=V_1z_q'(\theta)-x_1 \text{ and } \dfrac{\partial \Pi_2}{\partial x_2}=V_2z_q'(\theta)-x_2 $$

and 

$$\dfrac{\partial^2 \Pi_1}{\partial x_1^2}=V_1(z_q''(\theta))-1<0\text{ and } \dfrac{\partial^2 \Pi_2}{\partial x_2^2}=-V_2(z_q''(\theta))-1<0 .$$

From Assumption \ref{ass:existence2}, $\Pi_i$ is concave in $x_i$ for any $x_j$. Since any pure-strategy Nash equilibrium must satisfy the first-order conditions
$$V_1z_q'(\theta)=x_1  \text{ and } V_2 z_q'(\theta)=x_2,$$
concavity of payoffs implies that the second-order conditions for maximization are satisfied at any solution to the first-order conditions. 
Now any solution to the two first-order conditions must satisfy the following:

$$\dfrac{x_1}{x_2}=\dfrac{V_1}{V_2}$$
and $$x_1-x_2=(V_1-V_2)z_q'(x_1-x_2).$$

Consider the equation $\theta-(V_1-V_2)z_q'(\theta)=0$. The derivative of the left hand side is $1-(V_1-V_2)z_q''(\theta)$ which is $\geq 0$ as long as $z_q''(\theta) \leq \frac{1}{V_1-V_2}$. From Assumption \ref{ass:existence2}, $z_q''(\theta) \leq \frac{1}{V_1}$ which implies $ z_q''(\theta) \leq \frac{1}{V_1-V_2} $. Thus, the left hand side is monotone increasing in $\theta$. Also observe that at $\theta=0$, the left hand side is $-(V_1-V_2)z_q'(0)<0$. Thus, there is a unique solution to the equation $\theta-(V_1-V_2)z_q'(\theta)=0$. Let $\beta(q)$ denote this unique solution to the equation 
$\theta-(V_1-V_2)z_q'(\theta)=0$. The uniqueness of $\beta(q)$ implies that there is a unique solution to the first-order conditions and it takes the form
$$x_1^*=\dfrac{\beta(q)V_1}{V_1-V_2}  \text{ and }   x_2^*=\dfrac{\beta(q)V_2}{V_1-V_2}.$$
Equivalently, we can write it as 
$$x_1^*=V_1z_q'(\beta(q))   \text{ and } x_2^*=V_2z_q'(\beta(q)).$$
where $\beta(q)$ is the unique solution to the equation $\theta=(V_1-V_2)z_q'(\theta)$.
\end{proof}

To prove the result, we show that there  is a unique solution to the first-order conditions \ref{foc1} and \ref{foc2} for the contest game defined by $x_1^*$ and $x_2^*$. The concavity of the payoffs implied by Assumption \ref{ass:existence} means that the second-order conditions for maximization are satisfied\footnote{Assumption \ref{ass:existence2} implies that the players payoffs $\Pi_i$ are concave in $x_i$. The result goes through if the payoffs $\Pi_i$ are instead assumed to be quasiconcave in $x_i$.}. The full proof is in the appendix.

Note that the solution $\beta(q)$ to the equation 
$\theta=(V_1-V_2)z_q'(\theta)$ is unique and $\geq 0$ for all $q\in [0,1]$. It represents the difference in equilibrium effort levels exerted by the two agents. That is, the equilibrium effort levels $x_1^*, x_2^*$ are such that  $\frac{x_1^*}{x_2^*}=\frac{V_1}{V_2}$ and $x_1^*-x_2^*=\beta(q)$.

Next, we discuss the effect of the tie-breaking rule $q$ on total equilibrium effort.

\begin{thmrep}
\label{thm:diff}
Consider a difference-form contest satisfying Assumptions \ref{ass:existence2} and \ref{ass:tie2}. 

\begin{itemize}
   \item If $V_1>V_2$, $x_1^*(q)+x_2^*(q)$ is decreasing in $q$ and so the optimal tie-breaking rule is $q=0$. 
\item If $V_1=V_2$, $x_1^*(q)+x_2^*(q)$ is independent of $q$. 
\end{itemize}
\end{thmrep}

\begin{proof}[Proof of Theorem \ref{thm:diff}]
From Theorem \ref{thm:diff1}, the total effort in the unique pure-strategy Nash equilibrium is given by 
$$R(q)=(V_1+V_2)z_q'(\beta(q)),$$
where $\beta(q)$ is the unique solution to the equation $\theta=(V_1-V_2)z_q'(\theta).$

Consider the case where $V_1>V_2$. In this case, we have that $$R(q)=(V_1+V_2)z_q'(\beta(q))=\dfrac{V_1+V_2}{V_1-V_2}\beta(q)$$ from the equation defining $\beta(q)$. Thus,  $R'(q)=\dfrac{V_1+V_2}{V_1-V_2}\beta'(q)$ which implies that the total equilibrium effort $R(q)$ goes in the same direction as the difference in equilibrium effort $\beta(q)$ as we change $q$. From the characterizing equation, we know that
$$\beta'(q)=(V_1-V_2)\left(p''(\beta(q))\beta'(q)+p_0'(\beta(q))+qp_0''(\beta(q))\beta'(q)\right),$$
which implies 
$$\beta'(q)\left(1-(V_1-V_2)\left(p''(\beta(q))+qp_0''(\beta(q))\right)\right)=(V_1-V_2)p_0'(\beta(q)).$$

Given that $z_q''(\theta) \in [-\frac{1}{V_1},\frac{1}{V_1}]$ from Assumption \ref{ass:existence2}, we know that $\beta'(q)$ has the same sign as $p_0'(\beta(q))$. We already know $\beta(q)>0$ and therefore, from Assumption \ref{ass:tie2}, we get that $\beta'(q)<0$. Therefore, the total equilibrium effort $R(q)$ is decreasing in $q$ and it follows that the optimal tie breaking rule breaks ties in favor of the weaker agent by setting $q=0$. 

When $V_1=V_2$, $\beta(q)=0$ for all $q \in [0,1]$ and thus, $R(q)=(V_1+V_2)(p'(0)+qp_0'(0))$. But we know from Assumption \ref{ass:tie2} that $p_0'(0)=0$ and thus, we get that the total effort equals $R(q)=(V_1+V_2)p'(0)$ for any $q \in [0,1]$. Thus, with symmetric agents, the choice of the tie-breaking rule does not matter for equilibrium, and thus, total effort.
\end{proof}

To prove Theorem \ref{thm:diff}, we can use Theorem \ref{thm:diff1} and our assumptions on $p_0$ to see how the total effort changes as we increase $q$. When $V_1>V_2$, player $1$ puts in greater effort so that $p_0'(x_1^*-x_2^*)<0$. This implies that $R(q)$ is decreasing in $q$ and thus, the optimal tie-breaking rule sets $q=0$. In comparison, when $V_1=V_2$, the agents exert equal effort irrespective of $q$ and because $p_0'(0)=0$, it follows that the total effort does not depend on the choice of $q$. The full proof is in the appendix.\\



The difference-form contest success functions were first studied in \citet{hirshleifer1989conflict}. A well-known example, with zero probability of ties, is the logit function  $p_i(x_1,x_2)=\frac{\exp(x_i)}{\exp(x_i)+\exp(x_{-i})}$. While we are not aware of explicit work on generalizing these contest success functions to allow for the possibility of ties, we consider a couple of generalizations inspired from the literature on generalizing ratio-form contest success functions to allow for the possibility of ties.\\


The first generalization we consider is given by $p_i(x_i,x_j)=\left(\frac{\exp(x_i)}{\exp(x_i)+\exp(x_{-i})}\right)^k$ with $k \geq 1$.  In this case, we obtain the following lemma.

\begin{lemrep}
\label{lem:vesper2}
 Suppose $C$ is a difference-form contest with $p_i(x_i,x_j)=\left(\frac{\exp(x_i)}{\exp(x_i)+\exp(x_{-i})}\right)^k$ with $k \geq 1$. There exist some $\bar{V}>0$ such that if $\bar{V} \geq V_1 \geq V_2>0$, then for any $q\in [0,1]$, the contest has a unique pure-strategy Nash equilibrium defined by 
$$x_1^*=V_1z_q'(\beta(q))   \text{ and } x_2^*=V_2z_q'(\beta(q)),$$
where $\beta(q)$ is the unique solution to the equation $\theta=(V_1-V_2)z_q'(\theta)$.
\end{lemrep}

\begin{proof}
To prove the result, we define the functions $p(\theta), p_0(\theta)$ and show that they satisfy Assumptions \ref{ass:existence2}, \ref{ass:tie2} if $V_1$ is small enough. The result then follows from Theorems \ref{thm:diff1} and \ref{thm:diff}.

For the given contest, we can define the functions $$p(\theta)=\dfrac{e^{k\theta}}{(1+e^\theta)^k}, \quad p_0(\theta)=\dfrac{(1+e^\theta)^k-1-e^{k\theta}}{(1+e^\theta)^k}$$ 

so that $$z_q(\theta)=(1-q) \dfrac{e^{k\theta}}{(1+e^\theta)^k}+q\left(1-\dfrac{1}{(1+e^\theta)^k}\right).$$

Let us first identify conditions under which Assumption \ref{ass:existence2} is satisfied.

To do so, we have that
\begin{align*}
z_q'(\theta)&=\dfrac{ke^{\theta}}{(1+e^\theta)^{k+1}}\left((1-q)e^{(k-1)\theta}+q\right).
\end{align*}

Also, 
\begin{align*}
z_q''(\theta)&=\dfrac{ke^{\theta}}{(1+e^\theta)^{k+2}}\left((1-q)e^{(k-1)\theta}(k-e^\theta)+q(1-ke^\theta)\right).
\end{align*}

Observe that $z_q'(\theta)>0$. 
Letting $t=e^\theta$, we get
$$z_q''(\theta)=\dfrac{kt}{(1+t)^{k+2}}\left((1-q)t^{(k-1)}(k-t)+q(1-kt)\right).$$

Since $\lim_{t \to \infty} z_q''(\theta)=0$ and $\lim_{t \to 0} z_q''(\theta)=0$, there exists some bounds $m, M$ such that $m \leq z_q''(\theta) \leq M$ for all $\theta \in \mathbb{R}$. Thus, if we take $V_1$ small enough so that
$\frac{-1}{V_1} \leq m \leq M \leq \frac{1}{V_1}$, Assumption \ref{ass:existence2} will be satisfied. 

We also note that this contest success function satisfies Assumption \ref{ass:tie2} as 
$$p_0'(\theta)=\dfrac{ke^{\theta}}{(1+e^\theta)^{k+1}}(1-e^{(k-1)\theta}),$$
which is $>0$ for $\theta<0$ and $<0$ for $\theta>0$. Thus, Theorem \ref{thm:diff} applies.
\end{proof}

To prove the result, we show that $z_q'(\theta)>0$ and also $\lim_{\theta \to \infty} z_q''(\theta)=0$ and $\lim_{\theta \to -\infty} z_q''(\theta)=0$ which implies $z_q''(\theta)$ is bounded between some $M$ and $m$. Thus, if we take $V_1$ small enough so that
$\frac{-1}{V_1} \leq m \leq M \leq \frac{1}{V_1}$, Assumption \ref{ass:existence2} will be satisfied. We further show that this contest satisfies Assumption \ref{ass:tie2} so that Theorem \ref{thm:diff} applies.\\

The next generalization we consider is given by $p_i(x_i,x_j)=\frac{\exp(x_i)}{\exp(x_i)+k\exp(x_{-i})}$ with $k \geq 1$.  In this case, we obtain the following lemma.

\begin{lemrep}
\label{lem:jia2}
 Suppose $C$ is a difference-form contest with $p_i(x_i,x_j)=\frac{\exp(x_i)}{\exp(x_i)+k\exp(x_{-i})}$ with $k \geq 1$. There exist some $\bar{V}>0$ such that if $\bar{V} \geq V_1 \geq V_2>0$, then for any $q\in [0,1]$, the contest has a unique pure-strategy Nash equilibrium defined by 
$$x_1^*=V_1z_q'(\beta(q))   \text{ and } x_2^*=V_2z_q'(\beta(q)),$$
where $\beta(q)$ is the unique solution to the equation $\theta=(V_1-V_2)z_q'(\theta)$.
\end{lemrep}

\begin{proof}
To prove the result, we define the functions $p(\theta), p_0(\theta)$ and show that they satisfy Assumptions \ref{ass:existence2}, \ref{ass:tie2} if $V_1$ is small enough. The result then follows from Theorems \ref{thm:diff1} and \ref{thm:diff}.

For the given contest, we can define the functions $$p(\theta)=\dfrac{e^{\theta}}{k+e^\theta}, \quad p_0(\theta)=1-\dfrac{e^{\theta}}{k+e^\theta}-\dfrac{1}{ke^\theta+1} $$ 

so that $$z_q(\theta)=(1-q) \dfrac{e^{\theta}}{k+e^\theta}+q\dfrac{ke^\theta}{1+ke^\theta}.$$

Let us first identify conditions under which Assumption \ref{ass:existence2} is satisfied.

To do so, we have that

\begin{align*}
z_q'(\theta)&=ke^{\theta}\left(\dfrac{1-q}{(k+e^\theta)^2}+\dfrac{q}{(1+ke^\theta)^2}\right).
\end{align*}

Also, 

\begin{align*}
z_q''(\theta)&=ke^{\theta}\left(\dfrac{(1-q)(k^2-e^{2\theta})}{(k+e^\theta)^4}+\dfrac{q(1-k^2e^{2\theta})}{(1+ke^\theta)^4}\right).
\end{align*}

Again, we have that $z_q'(\theta)>0$ and also $\lim_{\theta \to \infty} z_q''(\theta)=0$ and $\lim_{\theta \to -\infty} z_q''(\theta)=0$ which implies $z_q''(\theta)$ is bounded between some $M$ and $m$. Thus, if we take $V_1$ small enough so that
$\frac{-1}{V_1} \leq m \leq M \leq \frac{1}{V_1}$, Assumption \ref{ass:existence2} will be satisfied.

We also note that this contest success function satisfies Assumption \ref{ass:tie2} as 
$$p_0'(\theta)=ke^{\theta}\left(\dfrac{1}{(1+ke^\theta)^2}-\dfrac{1}{(k+e^\theta)^2}\right),$$
which is $>0$ for $\theta<0$ and $<0$ for $\theta>0$. Thus, Theorem \ref{thm:diff} applies.
\end{proof}

As before, we show that $z_q'(\theta)>0$ and also $\lim_{\theta \to \infty} z_q''(\theta)=0$ and $\lim_{\theta \to -\infty} z_q''(\theta)=0$ which implies $z_q''(\theta)$ is bounded between some $M$ and $m$. It follows that  \ref{ass:existence2} and \ref{ass:tie2} will be satisfied if $V_1$ is small enough and thus, we can apply Theorems \ref{thm:diff1} and \ref{thm:diff} to get the result.\\

\section{Discussion}

In this section, we consider some other examples of contests and also discuss the design of random tie-breaking rules.

\subsection{Concave contests}

Here we consider contests $C=(V_1,V_2,p_1(),p_2(),q,c())$ that satisfy the following assumption.

\begin{assumption}
\label{ass:concave}

The contest success functions take the form $$p_i(x_1,x_2)=\dfrac{f_i(x_i)}{f_1(x_1)+f_2(x_2)+1}  \text{ and }  p_0(x_1,x_2)=\dfrac{1}{f_1(x_1)+f_2(x_2)+1}$$ where $f_i:\mathbb{R}_+\to \mathbb{R}_+ $ is twice-differentiable, strictly increasing, and concave.
\end{assumption}

This class of contest success functions with a possibility of tie was axiomatized by \citet{blavatskyy2010contest}. For such contest success functions, we have the following lemma.

\begin{lemrep}
Suppose the contest $C$ satisfies Assumption \ref{ass:concave}. Then, for any increasing and convex cost function $c$ with $c(0)=0$ and any tie-breaking rule $q\in [0,1]$,  the contest admits a unique pure-strategy Nash equilibrium.
\end{lemrep}
\begin{proof}
Let $f_1,f_2$ denote the increasing and concave impact functions that define the contest success functions $p_1,p_2,p_0$. 
For any $q \in [0,1]$, we can write the probability that player $i$ eventually wins the contest as $$P_i=\dfrac{g_i(x_i)}{g_1(x_1)+g_2(x_2)},$$ where $$g_1(x_1)=f_1(x_1)+qs  \text{ and }  g_2(x_2)=f_2(x_2)+(1-q)s.$$
Observe that $g_i$ is also twice-differentiable, strictly increasing and concave. Thus, it follows from Theorem 1 of \citet{fu2020optimal} that there exists a unique pure-strategy Nash equilibrium in this game.
\end{proof}

Given a tie-breaking rule $q \in [0,1]$, we can define impact functions $g_1(x_1)=f_1(x_1)+q$ and $g_2(x_2)=f_2(x_2)+(1-q)$ which, together with a convex cost function, constitute a regular concave contest (as defined in Definition 1 of \citet{fu2020optimal}). The result then follows from Theorem 1 of \citet{fu2020optimal} which says that there exists a unique pure-strategy Nash equilibrium in regular concave contest games\footnote{Observe that a tie-breaking rule essentially defines additive head-starts $q$ and $1-q$ in the \citet{blavatskyy2010contest} model. While previous work has studied the problem of finding optimal head-starts (\citet{franke2018optimal, fu2020optimal, li2023optimally}), our problem of finding the optimal tie-breaking rule is different in that the head-starts must add up to some constant.}. \\
The unique pure-strategy Nash equilibrium may involve both players exerting positive effort, in which case it is characterized by the first-order conditions \ref{foc1} and \ref{foc2}. Going forward, we will focus on the special case where $f_1(x)=f_2(x)=f(x)=x^r$ for $r \in (0,1]$ and $c(x)=x$. With these parametric assumptions, the first-order conditions are given by

$$V_1P_2rx_1^{r-1}=x_1^r+x_2^r+1=V_2P_1rx_2^{r-1},$$ where $P_2=p_2+(1-q)p_0$ and $P_1=p_1+qp_0$. 

We note here that for $r=1$, the solution to these conditions is
$$
x_1^*=\frac{V_1^2 V_2}{\left(V_1+V_2\right)^2}-q  \text{ and }  x_2^*=\frac{V_1 V_2^2}{\left(V_1+V_2\right)^2}-(1-q),
$$
and it will constitute the unique pure-strategy Nash equilibrium as long as $x_1^*, x_2^* \geq 0$. Observe that the headstarts defined by the tie-breaking rule $q$ act as a direct substitute for the effort exerted by them and the total equilibrium effort is independent of the choice of $q$. The next lemma illustrates how the tie-breaking rule matters in the case where $r=0.5$.

\begin{lemrep}
\label{blava}
Suppose the contest $C$ is such that $p_i(x_1,x_2)=\dfrac{\sqrt{x_i}}{\sqrt{x_i}+\sqrt{x_j}+1}$, $c(x)=x$ and $V_1=V_2=V$. 
\begin{itemize}
\item For any tie-breaking rule $q \in [0,1]$, the contest has a unique pure-strategy Nash equilibrium defined by $$x_1^*=\frac{1}{4(2V+1)}\left(V-(\sqrt{2V+1}-1)q\right)^2$$
and $$x_2^*=\frac{1}{4(2V+1)}\left(V-(\sqrt{2V+1}-1)(1-q)\right)^2.$$
\item The total effort $x_1^*+x_2^*$ is convex in $q \in [0,1]$ with a minimum at $q=0.5$ and maximum at $q=0$ and $q=1$.  
\end{itemize}
\end{lemrep}

\begin{proof}
For the general case of $r<1$ and $V_1\geq V_2$, we get from the first-order conditions that the equilibrium effort levels $x_1^*$ and $x_2^*$  satisfies the equations
$$ V_2x_1\left(1+\frac{q}{x_1^r}\right)=V_1x_2\left(1+\frac{1-q}{x_2^r}\right)$$ and

$$V_2r\left(x_1^r+q\right)=x_2^{1-r}\left(x_1^{r}+x_2^{r}+1\right)^{2}.$$

In this special case where $r=0.5$ with $V_1=V_2=V$, the equations simplify to:
$$ \sqrt{x_1}\left(\sqrt{x_1}+q\right)=\sqrt{x_2}\left(\sqrt{x_2}+1-q\right) \quad V\left(\sqrt{x_1}+q\right)=2\sqrt{x_2}\left(\sqrt{x_1}+\sqrt{x_2}+1\right)^{2}$$

We can solve the two equations to get the unique pure strategy Nash equilibrium as in the lemma. 

Let us verify that the $x_1^*$ and $x_2^*$ satisfy these two equations. To verify, let $a=V$ and $b=\sqrt{2V+1}$ so that $x_1^*=\left(\frac{1}{2b}(a-(b-1)q)\right)^2$ and $x_2^*=\left(\frac{1}{2b}(a-(b-1)(1-q))\right)^2$. 
Then we have that $$\sqrt{x_1^*}=\frac{1}{2b}\left(a-(b-1)q\right) \quad \sqrt{x_1^*}+q=\frac{1}{2b}\left(a+(b+1)q\right)$$ and
$$\sqrt{x_2^*}=\frac{1}{2b}\left(a-(b-1)(1-q)\right) \quad \sqrt{x_2^*}+1-q=\frac{1}{2b}\left(a+(b+1)(1-q)\right).$$ 
Using these, together we the fact that $2a=b^2-1$, we can verify that $x_1^*$ and $x_2^*$ solve the two first order conditions.

For the second part, observe that the total effort takes the form
$x_1^*+x_2^*=(c-dq)^2+(c-d(1-q))^2$ for appropriately defined constants $c$ and $d$ that depend only on $V$. Upon simplifying, the total effort equals $2c^2+d^2-2cd+2d^2q(q-1)$ which is clearly convex in $q$ and has a minimum at $q=0.5$ and maximum at $q=0,1$. 



\end{proof}

Even though the agents are identical, a designer who cares about maximizing effort is better off biasing the contest in favor of one of the agents by committing to breaking ties in its favor. Note that this is in contrast to ratio-form and difference-form contests where we saw that  with identical agents, the choice of tie-breaking rule has no effect on effort. 



\subsection{Random tie-breaking rules}

As we have discussed, there are many situations where a designer would want to bias the contest in favor of one of the players, even when they are ex-ante identical (as in Lemma \ref{blava}). But designing such a biased contest may be controversial or even infeasible if the designer is not aware of the relative strengths of the two players. In such situations, the designer may still be able to introduce bias into the contest while being fair ex-ante. For example, instead of arbitrarily committing to breaking ties in favor of one of the players, it can publicly toss a fair coin before the contest begins to pre-determine a winner in case the contest ends in a tie\footnote{Many sports have coin tosses before the contest begins which may introduce some bias. For example, in cricket, a coin is tossed before the match begins and the winner of the toss gets to choose whether they want to bat or bowl first. There is some \href{https://en.wikipedia.org/wiki/Toss_(cricket)}{evidence} that winning the toss provides a small, but significant improvement to a team's chances of winning.}. This corresponds to a \textit{random tie-breaking rule} in which the tie-breaking rule $q$ is $0$ or $1$ with equal probability. Note that this is different from the tie-breaking rule $q=0.5$ in which case a fair coin is tossed to determine the winner after the contest has ended in a tie.

More generally, a \textit{random tie-breaking rule} is defined by a random variable $Q$ with support in $[0,1]$ and we say that the rule is \textit{unbiased} if  $\mathbb{E}[Q]=0.5$. The timing of the contest with random tie-breaking rules is as follows:
\begin{enumerate}
    \item The contest designer chooses a distribution for random variable $Q$ with support in $[0,1]$. This is the random tie-breaking rule and is known to all participants.
    \item The value of $Q$ is realized and publicly revealed. This is the tie-breaking rule $q \in [0,1]$.
    \item Agents decide their effort levels.
    \item The contest outcome is revealed and in case of a tie, player 1 is chosen as winner with probability $q$ and player 2 is chosen as winner with probability $1-q$. 
\end{enumerate}

Next, we will discuss the design of random unbiased tie-breaking rules. More precisely, we will consider the problem of choosing a random tie-breaking rule $Q$ to maximize expected total effort $\mathbb{E}[x_1(Q)+x_2(Q)]$ under the constraint that the rule is unbiased so that $\mathbb{E}[Q]=0.5$.

First, for ratio-form contests, we obtain the following as a corollary of Theorem \ref{baik}.

\begin{corollary}
In a ratio-form contest satisfying Assumptions \ref{ass:existence} and \ref{ass:tie}, any random unbiased tie-breaking rule leads to the same expected total effort.
\end{corollary}

This follows from the fact that the total equilibrium effort is linear in $q$. 

For the difference-form contests, we identify a sufficient condition under which breaking ties by tossing a fair coin before the contest would be optimal. 

\begin{lemrep}
Consider a difference-form contest satisfying Assumptions \ref{ass:existence2} and \ref{ass:tie2}. If $V_1>V_2$ and $p_0''(\theta)<0$ for all $\theta \in [0, \sqrt{2V_1}]$, the total effort is convex in $q$ and the optimal unbiased random tie-breaking rule chooses $Q=0$ and $Q=1$ with equal probability.
\end{lemrep}

\begin{proof}
From Theorem \ref{thm:diff1}, the total effort in the unique pure-strategy Nash equilibrium is given by 
$$R(q)=(V_1+V_2)z_q'(\beta(q)) =\dfrac{V_1+V_2}{V_1-V_2}\beta(q),$$
where $\beta(q)$ is the unique solution to the equation $\theta=(V_1-V_2)z_q'(\theta).$

Thus, $$R''(q)=(V_1+V_2)z_q'(\beta(q)) =\dfrac{V_1+V_2}{V_1-V_2}\beta''(q).$$
Note that the solution $\beta(q)$ of an implicit equation $F(\theta, q)=0$ is convex iff
$$
\frac{\partial F}{\partial q} \frac{\partial^2 F}{\partial \theta \partial q}-\frac{\partial F}{\partial \theta} \frac{\partial^2 F}{\partial q^2} \geq 0.
$$
In our case, $\beta(q)$ is the solution to the equation $\left(V_1-V_2\right)\left(p^{\prime}(\theta)+q p_0^{\prime}(\theta)\right)-\theta=0$.

Then, $\beta(q)$ is convex if and only if

$$(V_1-V_2)^2p_0'(\theta)p_0''(\theta) \geq 0.$$

We know that for $V_1>V_2$,  $\theta>0$ and so $p_0'(\theta)<0$. Therefore, we need $p_0''(\theta)<0$ for $\beta(q)$, and thus, $R(q)$ to be convex. 
Observe also that player $1$'s equilibrium effort $x_1^*$ must be such that $V_1-\frac{x_1^2}{2} \geq 0 \iff x_1^* \leq \sqrt{2V_1}$. Since $x_2^*\geq 0$, it follows that $\beta(q) \leq \sqrt{2V_1}$ for any $q$. Thus, if we have $p_0''(\theta)\leq 0$ for $\theta \in [0, \sqrt{2V_1}]$, we have that the total effort is convex in $q$. It follows from convexity then that the optimal random tie-breaking rule in the class of unbiased rules breaks ties in fair manner before the contest.
\end{proof}

We note that when $V_1$ is small enough, the assumption $p_0''(\theta)<0$ for all $\theta \in [0, \sqrt{2V_1}]$ is satisfied for the difference-form contest success functions considered in Lemmas \ref{lem:vesper2} and \ref{lem:jia2}. 

Finally, for concave contests with square-root impact function and identical agents, it follows from Lemma \ref{blava} that breaking ties in a fair way before the contest maximizes effort among all random unbiased tie-breaking rules. Thus, a designer may be better off pre-determining the winner in case of a tie by tossing a fair coin before the contest begins as compared to the traditional practice of tossing a fair coin after the contest ends.

\section{Conclusion}

We study two-player contests with the possibility of ties under both ratio-form and difference-form contest success functions. In these contests, we study the effect different tie-breaking rules have on the effort exerted by the players. When players are heterogeneous, we find that the total effort decreases as the probability that ties are broken in favor of the stronger agent increases. Thus, an effort-maximizing designer would prefer to commit to breaking ties in favor of the weaker agent. The result lends further support to the encouraging effect of leveling the playing the field on effort in contests with heterogeneous agents. 

With symmetric agents, we find that the equilibrium is generally symmetric and does not depend on the choice of tie-breaking rule in the case of ratio-form and difference-form contests. The problem is more interesting for concave contests in which we make some parametric assumptions and find that an effort-maximizing designer would prefer to pre-determine the winner in case of a tie by tossing a fair coin before the contest begins as compared to the standard practice of breaking ties after the contest ends. We believe that the study of tie-breaking rules for concave contests and contests with more than two agents provide interesting directions for future research.

\nocite{*}
\newpage
\bibliographystyle{ecta}

\bibliography{refs}

\begin{thebibliography}{38}
\newcommand{\enquote}[1]{``#1''}
\expandafter\ifx\csname natexlab\endcsname\relax\def\natexlab#1{#1}\fi

\bibitem[\protect\citeauthoryear{Baik}{Baik}{1998}]{baik1998difference}
\textsc{Baik, K.~H.} (1998): \enquote{Difference-form contest success functions
  and effort levels in contests,} \emph{European Journal of Political Economy},
  14, 685--701.

\bibitem[\protect\citeauthoryear{Baik}{Baik}{2004}]{baik2004two}
---\hspace{-.1pt}---\hspace{-.1pt}--- (2004): \enquote{Two-player asymmetric
  contests with ratio-form contest success functions,} \emph{Economic Inquiry},
  42, 679--689.

\bibitem[\protect\citeauthoryear{Bastani, Giebe, and G{\"u}rtler}{Bastani
  et~al.}{2022}]{bastani2022simple}
\textsc{Bastani, S., T.~Giebe, and O.~G{\"u}rtler} (2022): \enquote{Simple
  equilibria in general contests,} \emph{Games and Economic Behavior}, 134,
  264--280.

\bibitem[\protect\citeauthoryear{Bevi{\'a} and Corch{\'o}n}{Bevi{\'a} and
  Corch{\'o}n}{2015}]{bevia2015relative}
\textsc{Bevi{\'a}, C. and L.~C. Corch{\'o}n} (2015): \enquote{Relative
  difference contest success function,} \emph{Theory and Decision}, 78,
  377--398.

\bibitem[\protect\citeauthoryear{Blavatskyy}{Blavatskyy}{2010}]{blavatskyy2010contest}
\textsc{Blavatskyy, P.~R.} (2010): \enquote{Contest success function with the
  possibility of a draw: Axiomatization,} \emph{Journal of Mathematical
  Economics}, 46, 267--276.

\bibitem[\protect\citeauthoryear{Chang, Wang, and Lv}{Chang
  et~al.}{2023}]{chang2023analysis}
\textsc{Chang, R., Y.~Wang, and L.~Lv} (2023): \enquote{An analysis of group
  contests with the possibility of a draw,} \emph{Journal of Industrial and
  Management Optimization}, 19, 3952--3975.

\bibitem[\protect\citeauthoryear{Che and Gale}{Che and
  Gale}{2000}]{che2000difference}
\textsc{Che, Y.-K. and I.~Gale} (2000): \enquote{Difference-form contests and
  the robustness of all-pay auctions,} \emph{Games and Economic Behavior}, 30,
  22--43.

\bibitem[\protect\citeauthoryear{Chowdhury, Esteve-Gonz{\'a}lez, and
  Mukherjee}{Chowdhury et~al.}{2023}]{chowdhury2023heterogeneity}
\textsc{Chowdhury, S.~M., P.~Esteve-Gonz{\'a}lez, and A.~Mukherjee} (2023):
  \enquote{Heterogeneity, leveling the playing field, and affirmative action in
  contests,} \emph{Southern Economic Journal}, 89, 924--974.

\bibitem[\protect\citeauthoryear{Cohen and Sela}{Cohen and
  Sela}{2007}]{cohen2007contests}
\textsc{Cohen, C. and A.~Sela} (2007): \enquote{Contests with ties,} \emph{The
  BE Journal of Theoretical Economics}, 7.

\bibitem[\protect\citeauthoryear{Corch{\'o}n and Serena}{Corch{\'o}n and
  Serena}{2018}]{corchon2018contest}
\textsc{Corch{\'o}n, L.~C. and M.~Serena} (2018): \enquote{Contest theory,} in
  \emph{Handbook of Game Theory and Industrial Organization, Volume II}, Edward
  Elgar Publishing, 125--146.

\bibitem[\protect\citeauthoryear{Deng, Wang, and Wu}{Deng
  et~al.}{2018}]{deng2018incentives}
\textsc{Deng, S., X.~Wang, and Z.~Wu} (2018): \enquote{Incentives in lottery
  contests with draws,} \emph{Economics Letters}, 163, 1--5.

\bibitem[\protect\citeauthoryear{Drugov and Ryvkin}{Drugov and
  Ryvkin}{2017}]{drugov2017biased}
\textsc{Drugov, M. and D.~Ryvkin} (2017): \enquote{Biased contests for
  symmetric players,} \emph{Games and Economic Behavior}, 103, 116--144.

\bibitem[\protect\citeauthoryear{Eden et~al.}{Eden
  et~al.}{2006}]{eden2006optimal}
\textsc{Eden, M. et~al.} (2006): \enquote{Optimal ties in contests,} Tech. rep.

\bibitem[\protect\citeauthoryear{Ewerhart}{Ewerhart}{2017}]{ewerhart2017revenue}
\textsc{Ewerhart, C.} (2017): \enquote{Revenue ranking of optimally biased
  contests: The case of two players,} \emph{Economics Letters}, 157, 167--170.

\bibitem[\protect\citeauthoryear{Franke, Leininger, and Wasser}{Franke
  et~al.}{2018}]{franke2018optimal}
\textsc{Franke, J., W.~Leininger, and C.~Wasser} (2018): \enquote{Optimal
  favoritism in all-pay auctions and lottery contests,} \emph{European Economic
  Review}, 104, 22--37.

\bibitem[\protect\citeauthoryear{Fu and Wu}{Fu and Wu}{2020}]{fu2020optimal}
\textsc{Fu, Q. and Z.~Wu} (2020): \enquote{On the optimal design of biased
  contests,} \emph{Theoretical Economics}, 15, 1435--1470.

\bibitem[\protect\citeauthoryear{Garfinkel and Skaperdas}{Garfinkel and
  Skaperdas}{2007}]{garfinkel2007economics}
\textsc{Garfinkel, M.~R. and S.~Skaperdas} (2007): \enquote{Economics of
  conflict: An overview,} \emph{Handbook of defense economics}, 2, 649--709.

\bibitem[\protect\citeauthoryear{Gelder, Kovenock, and Roberson}{Gelder
  et~al.}{2019}]{gelder2019all}
\textsc{Gelder, A., D.~Kovenock, and B.~Roberson} (2019): \enquote{All-pay
  auctions with ties,} \emph{Economic Theory}, 1--49.

\bibitem[\protect\citeauthoryear{Hirshleifer}{Hirshleifer}{1989}]{hirshleifer1989conflict}
\textsc{Hirshleifer, J.} (1989): \enquote{Conflict and rent-seeking success
  functions: Ratio vs. difference models of relative success,} \emph{Public
  choice}, 63, 101--112.

\bibitem[\protect\citeauthoryear{Imhof and Kr{\"a}kel}{Imhof and
  Kr{\"a}kel}{2014}]{imhof2014tournaments}
\textsc{Imhof, L. and M.~Kr{\"a}kel} (2014): \enquote{Tournaments with gaps,}
  \emph{Economics Letters}, 122, 211--214.

\bibitem[\protect\citeauthoryear{Imhof and Kr{\"a}kel}{Imhof and
  Kr{\"a}kel}{2016}]{imhof2016ex}
---\hspace{-.1pt}---\hspace{-.1pt}--- (2016): \enquote{Ex post unbalanced
  tournaments,} \emph{The RAND Journal of Economics}, 47, 73--98.

\bibitem[\protect\citeauthoryear{Jia}{Jia}{2012}]{jia2012contests}
\textsc{Jia, H.} (2012): \enquote{Contests with the probability of a draw: a
  stochastic foundation,} \emph{Economic record}, 88, 391--406.

\bibitem[\protect\citeauthoryear{Jia, Skaperdas, and Vaidya}{Jia
  et~al.}{2013}]{jia2013contest}
\textsc{Jia, H., S.~Skaperdas, and S.~Vaidya} (2013): \enquote{Contest
  functions: Theoretical foundations and issues in estimation,}
  \emph{International Journal of Industrial Organization}, 31, 211--222.

\bibitem[\protect\citeauthoryear{Li, Wu, and Xing}{Li
  et~al.}{2023}]{li2023optimally}
\textsc{Li, B., Z.~Wu, and Z.~Xing} (2023): \enquote{Optimally biased contests
  with draws,} \emph{Economics Letters}, 111076.

\bibitem[\protect\citeauthoryear{Loury}{Loury}{1979}]{loury1979market}
\textsc{Loury, G.~C.} (1979): \enquote{Market structure and innovation,}
  \emph{The quarterly journal of economics}, 93, 395--410.

\bibitem[\protect\citeauthoryear{Malueg and Yates}{Malueg and
  Yates}{2006}]{malueg2006equilibria}
\textsc{Malueg, D.~A. and A.~J. Yates} (2006): \enquote{Equilibria in
  rent-seeking contests with homogeneous success functions,} \emph{Economic
  Theory}, 719--727.

\bibitem[\protect\citeauthoryear{Mealem and Nitzan}{Mealem and
  Nitzan}{2016}]{mealem2016discrimination}
\textsc{Mealem, Y. and S.~Nitzan} (2016): \enquote{Discrimination in contests:
  a survey,} \emph{Review of Economic Design}, 20, 145--172.

\bibitem[\protect\citeauthoryear{Minchuk}{Minchuk}{2022}]{minchuk2022winner}
\textsc{Minchuk, Y.} (2022): \enquote{Winner-pay contests with a no-winner
  possibility,} \emph{Managerial and Decision Economics}, 43, 1874--1879.

\bibitem[\protect\citeauthoryear{Nalebuff and Stiglitz}{Nalebuff and
  Stiglitz}{1983}]{nalebuff1983prizes}
\textsc{Nalebuff, B.~J. and J.~E. Stiglitz} (1983): \enquote{Prizes and
  incentives: towards a general theory of compensation and competition,}
  \emph{The Bell Journal of Economics}, 21--43.

\bibitem[\protect\citeauthoryear{Nti}{Nti}{1997}]{nti1997comparative}
\textsc{Nti, K.~O.} (1997): \enquote{Comparative statics of contests and
  rent-seeking games,} \emph{International Economic Review}, 43--59.

\bibitem[\protect\citeauthoryear{Nti}{Nti}{1999}]{nti1999rent}
---\hspace{-.1pt}---\hspace{-.1pt}--- (1999): \enquote{Rent-seeking with
  asymmetric valuations,} \emph{Public Choice}, 98, 415--430.

\bibitem[\protect\citeauthoryear{Nti}{Nti}{2004}]{nti2004maximum}
---\hspace{-.1pt}---\hspace{-.1pt}--- (2004): \enquote{Maximum efforts in
  contests with asymmetric valuations,} \emph{European journal of political
  economy}, 20, 1059--1066.

\bibitem[\protect\citeauthoryear{Skaperdas}{Skaperdas}{1996}]{skaperdas1996contest}
\textsc{Skaperdas, S.} (1996): \enquote{Contest success functions,}
  \emph{Economic theory}, 7, 283--290.

\bibitem[\protect\citeauthoryear{Szech}{Szech}{2015}]{szech2015tie}
\textsc{Szech, N.} (2015): \enquote{Tie-breaks and bid-caps in all-pay
  auctions,} \emph{Games and Economic Behavior}, 92, 138--149.

\bibitem[\protect\citeauthoryear{Tullock}{Tullock}{2001}]{tullock2001efficient}
\textsc{Tullock, G.} (2001): \enquote{Efficient rent seeking,} \emph{Efficient
  rent-seeking: Chronicle of an intellectual quagmire}, 3--16.

\bibitem[\protect\citeauthoryear{Vesperoni and Yildizparlak}{Vesperoni and
  Yildizparlak}{2019}]{vesperoni2019contests}
\textsc{Vesperoni, A. and A.~Yildizparlak} (2019): \enquote{Contests with
  draws: Axiomatization and equilibrium,} \emph{Economic inquiry}, 57,
  1597--1616.

\bibitem[\protect\citeauthoryear{Wang}{Wang}{2010}]{wang2010optimal}
\textsc{Wang, Z.} (2010): \enquote{The optimal accuracy level in asymmetric
  contests,} \emph{The BE Journal of Theoretical Economics}, 10.

\bibitem[\protect\citeauthoryear{Zhu}{Zhu}{2021}]{zhu2021optimal}
\textsc{Zhu, F.} (2021): \enquote{On optimal favoritism in all-pay contests,}
  \emph{Journal of Mathematical Economics}, 95, 102472.

\end{thebibliography}

\end{document}